# Signatures of the Optical Stark Effect on Entangled Photon Pairs from Resonantly-Pumped Quantum Dots

F. Basso Basset[1,*], M. B. Rota[1,*], M. Beccaceci[1], T. M. Krieger[2], Q. Buchinger[3], J. Neuwirth[1], H. Huet[1], S. Stroj[4], S. F. Covre da Silva[2], G. Ronco[1], C. Schimpf[2], S. Höfling[3], T. Huber-Loyola[3], A. Rastelli[2], and R. Trotta[1,†]

[1]*Department of Physics, Sapienza University of Rome, 00185 Rome, Italy*
[2]*Institute of Semiconductor and Solid State Physics, Johannes Kepler University, 4040 Linz, Austria*
[3]*Technical Physics, University of Würzburg, 97074 Würzburg, Germany*
[4]*Forschungszentrum Mikrotechnik, FH Vorarlberg, 6850 Dornbirn, Austria*

Two-photon resonant excitation of the biexciton-exciton cascade in a quantum dot generates highly polarization-entangled photon pairs in a near-deterministic way. However, the ultimate level of achievable entanglement is still debated. Here, we observe the impact of the laser-induced AC-Stark effect on the quantum dot emission spectra and on entanglement. For increasing pulse-duration/lifetime ratios and pump powers, decreasing values of concurrence are recorded. Nonetheless, additional contributions are still required to fully account for the observed below-unity concurrence.

The generation of entangled photons is essential to quantum communication [1–3], with relevance for multi-node connections [4] and quantum repeaters [5], information processing [6,7], imaging [8], and non-local networks [9,10]. As applications grow in complexity, increasingly efficient sources of highly entangled photons become paramount [11]. As an example, in a device-independent quantum key distribution protocol [12] a minimum concurrence—entanglement figure of merit, assuming a depolarizing channel—of 0.79 is required to achieve a positive secret key rate, which becomes higher the closer the concurrence is to unity. Moreover, some performance overhead is needed for long-distance communication due to the lower signal-to-noise ratio and below-unity fidelity of prospective quantum memories and repeater nodes.

This objective has spurred research on quantum dots (QDs), which produce pairs of polarization-entangled photons via the biexciton-exciton (XX-X) cascade [13]. Using two-photon resonant excitation (TPE), they provide nearly deterministic pair generation [14,15], with demonstrated probability above 90% [16,17], and a high degree of entanglement, with concurrence of 0.92–0.97 in a state-of-the-art system [18]. On-demand emission clocked at a fast and well-defined repetition rate is crucial for scalability, because the success rate of a protocol depends on the generation probability to the power of the number of photons involved and synchronization enables efficient coincidence-based operations. However, the reported values of entanglement, while enabling many applications, are still below what spontaneous parametric down-conversion in nonlinear crystals achieves [19], albeit at the cost of a photon pair generation probability which stays below a few percent [20].

A major research question is whether any limitation hinders the XX-X cascade from generating maximally entangled photon pairs, i.e., near-unity concurrence. A detrimental factor for as-grown QDs is the presence of a fine structure splitting (FSS) between the intermediate states of the radiative cascade, which introduces a phase evolution that makes the state mixed, unless time-gated. However, external tuning fields can be leveraged to remove this obstacle [21,22]. Another issue affecting first implementations based on above-band excitation is re-excitation [23–25]. This process happens when the system relaxes to the ground state fast enough that the pump still has a chance to drive it back into the excited state within a given temporal bin, resulting in multiphoton emission. Yet, TPE drastically reduces it using a resonant laser pulse much shorter that the time typically required to observe the full radiative cascade [26,27], leading to negligible impact on the two-photon state [28].

Entanglement is also resistant to common causes of dephasing in the solid state. Equal energy shifts in the bright X states do not reduce coherence in the two-photon state [29], factoring out many phonon-based processes [30–32]. Time-resolved experiments have shown that exciton cross-dephasing can be several times slower than the optical transition [33–35] and possibly ascribed to a coherent precession [36].

Nonetheless, Seidelmann et al. [37] recently predicted that TPE itself limits the degree of entanglement via the AC-Stark shift induced by the laser pulse. This phenomenon, also known as optical Stark effect, has been controllably induced by an additional near-resonant continuous-wave field on the XX and X states in a QD [38]. It has even been used to reduce the FSS to observe entanglement [39] but without resorting to a TPE scheme. These experiments focused on FSS control rather than on on-demand generation and investigated the effect as an external tuning knob but not as an intrinsic consequence of the optical excitation.

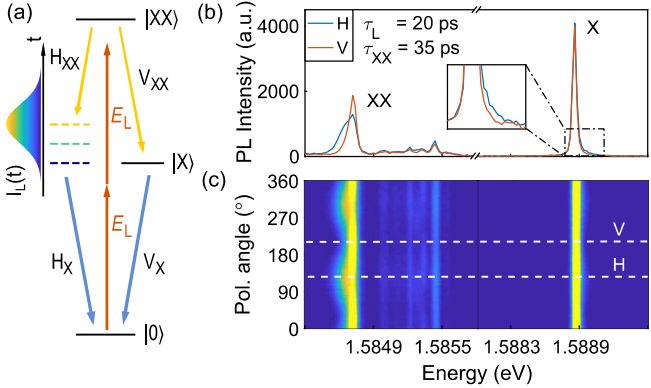

FIG. 1. (a) Energy diagram of the XX-X cascade under TPE. The transient blueshift of the X state aligned with the polarization of the laser (H) is shown with dotted energy levels at three times during the laser pulse, color-matched with the respective value of the laser intensity envelope function $I_L(t)$, sketched on the left. (b) Photoluminescence spectra of a QD under TPE at π-pulse area, selecting the linear polarization parallel and orthogonal (V) to the laser electric field. The spectral ranges are selected around the X and XX lines. (c) Extended data series, including intermediate linear polarizations. The spectra in (b) are marked by dashed lines.

In this work, we experimentally investigate for the first time how the optical Stark effect impacts entangled light emission from a single GaAs QD under different TPE conditions. The effect introduces a relevant performance tradeoff as the sample development naturally progresses towards faster operation rates. Accordingly, the study is performed on a novel device [40–42] able to provide desirable fast emission rates via the Purcell effect.

The process of entangled photon generation is illustrated in Fig. 1(a). Resonant TPE populates with high probability the XX state, formed by two confined electron-hole pairs. Assuming degenerate bright X states, the XX state decays radiating two photons in the state $|\phi^+\rangle = 1/\sqrt{2}\,(|HH\rangle + |VV\rangle)$, H/V being a basis of orthogonal linear polarizations.

Even with a symmetrical in-plane confinement potential, the degeneracy of the X levels is lifted by an external field. The linearly polarized laser pulse, tuned at half the energy of the XX state in TPE, marked as $E_L$ in the figure, blueshifts the X level with the dipole aligned to its electric field (H in our example). The shift depends on the laser intensity and causes an effective transient energy splitting of the intermediate state of the cascade, affecting the energy of the photons emitted during the laser pulse.

When the laser intensity is set to maximize population of the upper state, namely at the first maximum versus power in the Rabi cycles that describe the coherent excitation process—defined as π-pulse area [43]—, the effect is non-negligible. Its expected impact on entanglement is equivalent to that of a constant FSS of 200 μeV which is switched on for a typical laser pulse duration of 10 ps [37]. The energy difference adds a time-dependent phase to the two-photon polarization state for the fraction of photons emitted during the pulse. In the example, the effective splitting corresponds to a 21 ps precession period, which is short compared to the typical uncertainty in the emission time, leading to a degraded measured degree of entanglement [29]. Therefore, the entanglement critically depends on the ratio between pulse duration $\tau_L$ (FWHM of the intensity envelope function) and XX lifetime $\tau_{XX}$.

To observe the optical Stark effect, we control this variable up to where $\tau_L$ is comparable to $\tau_{XX}$. First, we design the excitation setup with variable laser pulse duration. A Ti:Sapphire femtosecond laser is equipped with a 4f pulse slicer to control the spectral bandwidth. The TPE bandwidth needs to be lower than the energy separation between the two emission lines—4 meV for the GaAs QDs studied here [44]—to avoid direct excitation of the X level. While we observe constant photoluminescence intensity for pulse widths up to 2.3 meV FWHM, corresponding to $\tau_L \approx 0.7(2)$ ps (see Supplemental Material), we stay below 1170 μeV to thoroughly filter the laser back-reflection. The spectral resolution of the pulse slicer sets the narrowest bandwidth to 78 μeV. The resulting pulse duration interval is 1.3(3)–20(4) ps, assuming a time-bandwidth product of 0.374. This was estimated on the laser before the pulse slicer, as reported in the Supplemental Material together with a sketch of the excitation setup and the laser spectra.

The second relevant quantity is the XX lifetime. We adopt GaAs QDs fabricated by droplet etching emitting in the 780–795 nm range. They offer high values of entanglement without temporal or spectral filtering [18] and natively feature fast emission, with typical $\tau_{XX}$ of 130 ps [17]. To reach transition lifetimes close to the laser pulse duration we embed single QDs in circular Bragg grating resonators [40,41], allowing for Purcell factors $F_p$ up to 3.7. The resonators are integrated onto a piezoelectric substrate to induce in-plane anisotropic strain and erase any FSS. Further details are provided in Ref. [42]. Such a device more closely approaches the operation targeted by the developments in high-rate entangled photon pair generation.

We first consider a QD featuring lifetimes of 35(1) ps and 53(1) ps for the XX and X states, respectively. Its FSS is strain-tuned down to 0.8(5) μeV, so to individually pinpoint the effect of the laser pulse. The AC-Stark shift induced under TPE appears in polarization-resolved photoluminescence, as illustrated in Fig. 1(b-c) for $\tau_L = 20(4)$ ps and π-pulse excitation. Two polarized sidebands are present on the low-energy side of the XX line (XX-X transition) and, weaker, on the high-energy side of the X (X-ground-state transition). Their polarization is parallel to that of the excitation laser and can be controlled by rotating its in-plane polarization angle (see Supplemental Material). Consistently with the optical Stark effect, the sideband is interpreted as integrating a laser-induced shift over different emission times following the lifetime distribution of the excitons. The effect is stronger on the XX emission, which takes place first in the cascade, is faster than the X emission, thus more likely to happen while the laser intensity on the QD is still non-negligible.

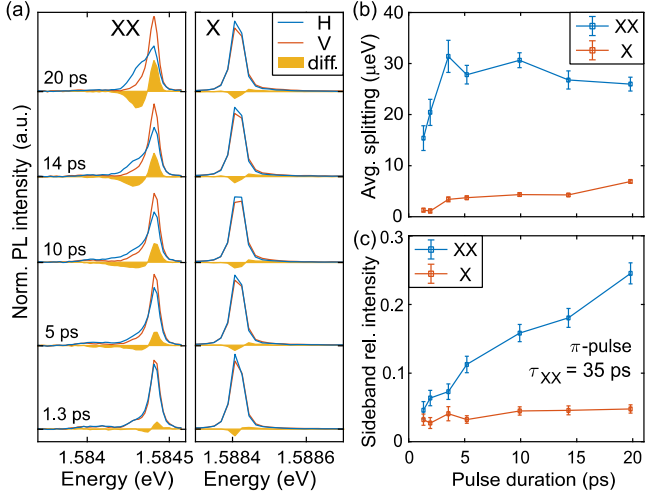

FIG. 2. (a) Photoluminescence spectra of a QD with $\tau_{XX} = 35(1)$ ps under TPE for fixed laser $\pi$-pulse area and different pulse durations. The X and XX lines are shown in the linear polarization parallel (H) and orthogonal (V) to the laser electric field, together with their difference. (b) Average energy splitting induced by the laser and (c) half-area of the difference between the V- and H-polarized spectra (whose area is normalized to one), from a spectral interval around the X and XX lines at $\pi$-pulse area.

Additional insight comes from varying the main excitation parameters. Figure 2(a) reports the XX and X lines, with their integrated intensity normalized to one, from the same QD at $\pi$-pulse area. The laser pulse duration is increased while the peak amplitude is reduced so that the average power is kept constant at the value which maximizes pumping into the XX state. Thus, the XX sideband appreciably grows in intensity along the H polarization, while the components orthogonal to the laser remain almost unchanged. Consistently, the polarized sideband is more pronounced also when the pump power is increased, as discussed in Appendix A.

Two quantities of interest are plotted in Fig. 2(b-c). Each point corresponds to a polarization-resolved measurement as in Fig. 1(c). The spectral peak position is calculated averaging the emission energies weighted by the corresponding photoluminescence intensity, in the 1.5882–1.5890 eV interval for the X line, in 1.5835–1.5846 eV for the XX. A small intensity offset is subtracted from the XX line as emission background. The series of spectral positions is then fitted with a sinusoidal function whose amplitude is the average energy splitting. Compared to the usual fit procedure for estimating the FSS [23], this method is more sensitive to noise but does not rely on any assumption on the peak shape, which results from a non-trivial weighted integral of time-dependent energy shifts.

The average XX photon energy splitting in Fig. 2(b) first raises and then saturates as a function of laser pulse duration. This is because the laser intensity is fixed at $\pi$-pulse area. If the pulse duration is increased, the peak intensity decreases accordingly. A smaller energy splitting is experienced by the QD for a longer temporal span, but its average over the lifetime distribution will not appreciably vary unless $\tau_L \ll \tau_{XX}$, in which case the chance of observing the AC-Stark shift becomes negligible, and the XX sideband is difficult to discriminate from additional emission lines.

Nonetheless, we expect that the pulse duration influences the entanglement even at fixed pulse area. The transient energy splitting present within the FWHM of the laser pulse is much larger than the radiative linewidth of the X state, even for the low intensities of longer pulses. This introduces a random phase that decreases the overall coherence terms of the measured two-photon density matrix. Therefore, a better figure of the impact of the optical Stark effect on entanglement is the fraction of photons emitted during the laser pulse. We approximate it by taking the difference between the V- and H-polarized emission spectra, after normalizing their area to one, and calculating the area below the positive values of the curve. A noise baseline is estimated from a portion of the spectra where no signal is present and subtracted. While this metric is affected by the spectrometer resolution, it provides a rough estimate of the photons emitted in the laser-induced sideband, for which the time evolution of the bright X states is significantly different from the decay events occurring after the laser pulse. Figure 2(c) shows that more photons are emitted in the XX sideband as $\tau_L$ gets closer to $\tau_{XX}$.

We now focus on the effect of these spectral features on entanglement. We collect data at $\pi$-pulse, for maximum photon flux, while varying the laser pulse duration. We perform quantum state tomography of the polarization of the XX and X photons using 36 coincidence measurements and a maximum likelihood estimation [45]. The concurrence extracted from the density matrix for the QD investigated so far is in filled orange rhombuses in Fig. 3(a), compared with the data in blue relative to a QD with a longer XX (X) lifetime of 81(3) ps (122(10) ps) and a FSS of 0.7(5) µeV.

The concurrence decreases for increasing pulse duration and, for a fixed laser pulse, is lower for faster XX recombination. Therefore, it depends on the ratio between $\tau_L$ and $\tau_{XX}$, according to the predictions of Ref. [37].

A known detrimental effect which depends on $\tau_L/\tau_{XX}$ is the multiphoton generation caused by re-excitation of the XX-X cascade, already mentioned in the introduction. We quantify it by measuring the zero-time delay autocorrelation function $g^{(2)}(0)$, as shown in Fig. 3(b-c), using a time window which includes detection events associated to the same laser pulse. Over a baseline attributed to background emission induced by a weak continuous-wave off-resonant light, added to photo-neutralize the QD [18,46], the $g^{(2)}(0)$ increases with pulse duration and more markedly for the QD with shorter lifetime. The values are compatible with theoretical expectations [27].

This level of multiphoton emission has limited impact on the concurrence, as shown by the circles in Fig. 3(a), which were simulated by subtracting its contribution using the measured $g^{(2)}(0)$ values [28].

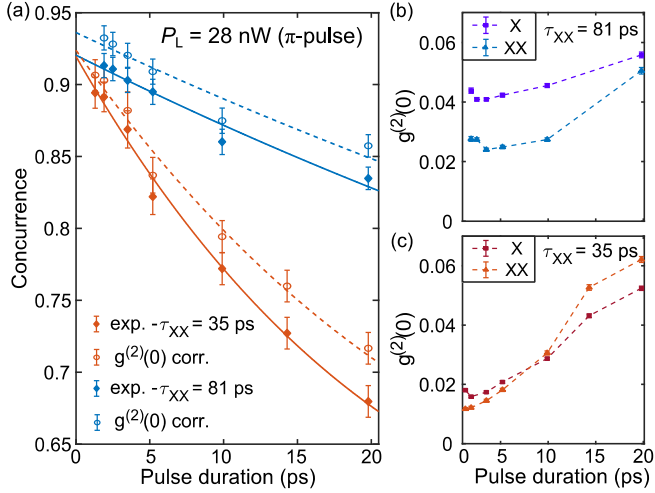

FIG. 3. (a) Concurrence of the photon pairs emitted from two QDs under TPE at π-pulse area for different laser pulse durations. The orange rhombuses refer to the QD with $\tau_{XX}$ = 35(1) ps ($F_p$ = 3.7), while the blue ones to another with $\tau_{XX}$ = 81(3) ps ($F_p$ = 1.6). The circles represent data corrected for multiphoton emission, estimated from the experimental $g^{(2)}(0)$ values. The lines are fits with Eq. (1). (b) Autocorrelation function at zero-time delay $g^{(2)}(0)$ for the X and XX lines of the QD with $\tau_{XX}$ = 81(3) ps. (c) Same as (b) for the QD with $\tau_{XX}$ = 35(1) ps.

The drop in concurrence with increasing pulse duration is mainly linked to the laser-induced AC-Stark shift instead. Its modeling in a dressed state picture leads to a complete description of the two-photon density matrix [37]. An effective model assuming an equivalent square laser pulse and negligible FSS gives a good analytical approximation for the concurrence:

$$C \approx C_0 \left(1 - \frac{\sqrt{2}\,\tau_L}{4\,\tau_{XX}} \exp\left(-\frac{\sqrt{2}\,\tau_L}{4\,\tau_{XX}}\right)\right) \quad (1)$$

where the factor $C_0$ accounts for other imperfections that limit entanglement independently from $\tau_L$.

Equation 1, with two free parameters, $C_0$ and $\tau_{XX}$, fits the concurrence drop as shown by the lines in Fig. 3, supporting its interpretation as consequence of the laser-induced energy splitting. The returned values for the XX lifetime are close to the measured ones, with $\tau_{XX}^{[fit]}$ = 22(5) ps for the QD with $\tau_{XX}$ = 35(1) ps, and $\tau_{XX}^{[fit]}$ = 68(19) ps for the QD with $\tau_{XX}$ = 81(3) ps. The accuracy may be limited by the simplified assumption on the time-bandwidth product. Phonon-induced dephasing may further decrease the concurrence in presence of an X splitting [32,47], but it is expected to be secondary for QDs of large size at 5 K temperature. In fact, the model reproduces the experimental dependence on the $\tau_L/\tau_{XX}$ ratio without considering features in the electronic structure beyond the four-level scheme or interactions with the solid-state environment.

The observed phenomenon introduces a potential tradeoff in designing a quantum emitter-based entangled photon source. On the one hand, the faster emission achieved via Purcell enhancement is advantageous for many reasons: increased clock rate of operation, minimized impact of dephasing mechanisms [18], and possibly more indistinguishable photons from a radiative cascade [48]. On the other hand, without changes to the excitation scheme an increased $\tau_L/\tau_{XX}$ ratio exacerbates the downsides of the laser-induced splitting. The laser pulse cannot be shortened indefinitely because its bandwidth must be lower that the energy difference between the two photons, namely the XX binding energy—a property dependent on the QD materials and geometry—to prevent direct excitation of the X state. In our case, photoluminescence intensity measurements indicate that shortening the laser pulse below 0.7 ps starts causing this detrimental process. This, in turn, places a lower limit on the emission time. For the higher Purcell enhancement considered ($\tau_{XX}$ = 35(1) ps), the optical Stark effect alone for the shorter pulse duration of 1.3 ps causes a 0.02 concurrence drop compared to the interpolated maximum at zero pulse duration.

These conflicting requirements limit the achievable clock rate at near-maximal entanglement levels in the range of several GHz. The ceiling falls even closer to the GHz threshold if aiming to accelerate the XX decay more than the X one to reduce temporal correlations in the radiative cascade and improve photon indistinguishability [48], although other strategies based on quantum interference may be viable depending on the application [49]. However, we can envisage possible mitigation approaches. As underlined through the spectral analysis, the lower entanglement is linked to a sideband emission, suggesting spectral filtering as a simple solution. While this implies a small deviation from on-demand generation, it could result in a reasonable tradeoff for most applications, similarly to what argued for QD single-photon sources [50]. We could demonstrate it in the worst-case scenario of long laser pulse and short recombination by removing part of the low-energy sideband of the XX line with a notch filter and partially increasing the concurrence, as reported in the Supplemental Material. For maximal effectiveness, a filter with high edge steepness and minimal spectral wandering on the emission line are needed. Alternatively, the search for efficient photon pair generation and maximal entanglement motivates research into original pumping schemes, such as solutions based on two-color excitation pulses [51]. These considerations also apply to photon indistinguishability, that is strongly affected by spectral line shifts.

We further notice from Fig. 3(a) that $C_0$, the expected concurrence at vanishing pulse duration, is below unity and amounts to 0.92(1) and 0.94(1) for the faster and slower QD respectively, after accounting for the estimated multiphoton component. If we also subtract the contribution from the residual FSS, for the QD with the shorter XX lifetime at best we obtain a concurrence of 0.93(1) and a maximal Bell-state fidelity of 0.96(1). Other physical mechanisms limiting the degree of entanglement are present, which have not been quantified yet. A role of spin noise and the Overhauser effect has been supported by theoretical modeling, spin coherence measurements, and lower figures of merit observed in In-based QDs [52,53]. However, its expected contribution is less than one percentage point of fidelity for the shortest emission times observed here, failing to explain the persisting gap. Further investigation is required to understand the missing relevant factors, such as a more nuanced modeling of the QD environment or neglected degrees of freedom, and devise optimized solutions.

In conclusion, we experimentally demonstrated the impact of the AC-Stark induced shift on the spectra and entanglement of photons emitted from the XX-X cascade under TPE. Using a QD source able to provide fast recombination and FSS suppression, we isolated the effect and investigated it down to a laser pulse duration comparable with the XX lifetime. We show that the optimal laser pulse results from the bandwidth requirements for TPE and minimizing the impact of the optical Stark effect. This imposes a tradeoff between the degree of entanglement and the source efficiency as faster emission is pursued to push the clock rate of operation beyond the GHz, yet it does not represent the ultimate limitation for the concurrence values observed in the current generation of devices. We anticipate that further research in this promising direction will lead to reliably design quantum emitters and excitation schemes capable of approaching the quantum limits for bipartite correlations.

This project has received funding from the European Research Council (ERC) under the European Union's Horizon 2020 Research and Innovation Program under Grant Agreement No. 679183 (SPQRel), by the European Union's Horizon 2020 Research and Innovation Program under Grant Agreements No. 899814 (Qurope), 871130 (Ascent+), within the QuantERA II Programme that has received funding from the European Union's Horizon 2020 research and innovation programme under Grant Agreement No. 101017733 via the project QD-E-QKD, from MUR (Ministero dell'Università e della Ricerca) through the PNRR MUR project PE0000023-NQSTI, from the FFG Grant No. 891366, from the Austrian Science Fund (FWF) via the Research Group FG5, P 29603, I 4380, I 3762, the Linz Institute of Technology (LIT), and the LIT Secure and Correct Systems Lab, supported by the State of Upper Austria. T.H.-L. acknowledges funding from the Federal Ministry of Education and Research (BMBF) through the Quantum Futur (FKZ: 13N16272) initiative.

*Appendix A: Power dependence.*—Another observation in support of interpreting the data relying on the optical Stark effect is the critical role of the pump power. We selected a series of values around the π-pulse area, corresponding to an average power on the QD of 28 nW, and report in Fig. 4(a-c) the analysis of the polarization-resolved spectra for the pulse duration of 20(4) ps, at which the effect is larger. The average energy splitting changes with the laser power, for the X line as well, even if it is much less affected compared to the XX. A close-up of the X spectra in Fig. 4(a) is in the Supplemental Material. Above π-pulse, the spectral shape of the XX line begins to be affected also in the polarization orthogonal to the one of the laser. This could be due to a different laser polarization direction passing from the collimated beam in the laboratory to the strongly focused beam in the optical cavity embedding the QD. Additionally, we point out that the energy splitting quickly varies close to π-pulse, where the emission intensity has a stationary point. From a practical perspective, it implies that trying to set the π-pulse condition by maximizing the brightness, but without monitoring the laser power, could allow for small deviations in power that result in significant variations of the laser-induced splitting, easily amounting to 20% in the example of Fig. 4(b).

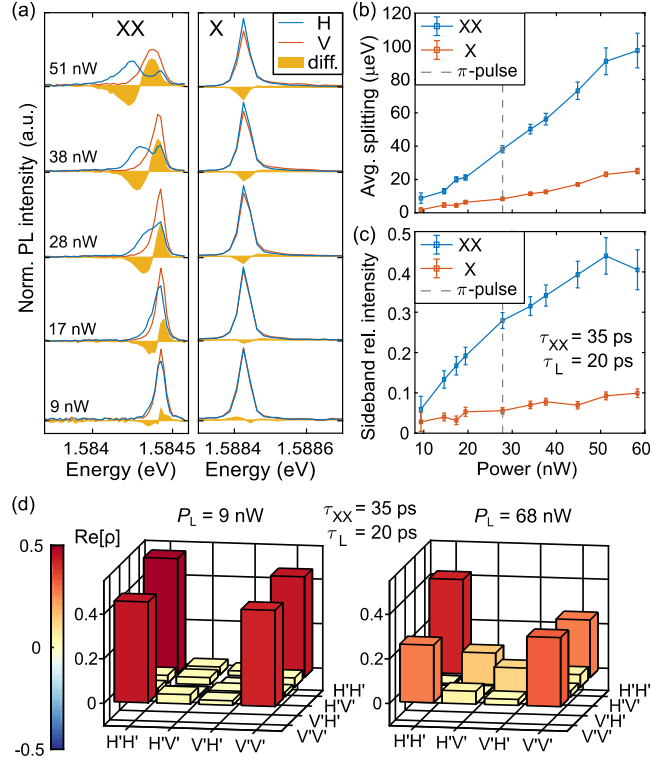

FIG. 4. (a) Photoluminescence spectra of a QD with $\tau_{XX} = 35(1)$ ps under TPE at different excitation powers, varying the pulse area around π-pulse while fixing the pulse duration at $\tau_L = 20(4)$ ps. The X and XX lines are shown in the linear polarization parallel (H) and orthogonal (V) to the laser electric field, together with their difference. (b) Average energy splitting induced by the laser and (c) half-area of the difference between the V- and H-polarized spectra (whose area is normalized to one), from a spectral interval around the X and XX line. (d) Real part of the polarization density matrices from XX-X coincidences for two different excitation powers $P_L$, below π-pulse and close to the first minimum in the Rabi cycles, that is 2π-pulse area.

Consistently with the optical Stark effect explanation, the degree of entanglement is strongly affected by laser power as well. In the conditions explored in Fig. 4(a-c), at minimum power the concurrence is 0.88(2), similar to what observed for the shorter laser pulses but at π-pulse. Instead, if the power approaches the first minimum in the Rabi cycles, that is 2π-pulse area, the concurrence falls to 0.27(1). The real part of the corresponding density matrices is shown in Fig. 4(d). The negligible imaginary terms, together with the matrix at π-pulse, are reported in the Supplemental Material.

---

*These authors contributed equally to this work.
†rinaldo.trotta@uniroma1.it